\documentclass[times,twocolumn,final]{elsarticle}

\usepackage{amsmath}
\usepackage{medima}
\usepackage{framed,multirow}

\usepackage{amssymb}
\usepackage{latexsym}

\usepackage{url}
\usepackage{xcolor}

\usepackage{hyperref}

\definecolor{newcolor}{rgb}{.8,.349,.1}

\begin{document}


\begin{frontmatter}

\title{B-Spine: Learning B-Spline Curve Representation for Robust and Interpretable Spinal Curvature Estimation}

\author[1]{Hao Wang}
\author[2]{Qiang Song}
\author[2]{Ruofeng Yin\corref{cor1}}
\author[1,3]{Rui Ma\corref{cor1}}
\author[4]{Yizhou Yu}
\author[1,3]{Yi Chang}
\cortext[cor1]{Corresponding authors. \newline
\indent \;\; \textit{Email addresses:} ruim@jlu.edu.cn (R. Ma), yrf\textunderscore wind@jlu.edu.cn (R. Yin).
}

\address[1]{School of Artificial Intelligence, Jilin University, Changchun, 130012, China}
\address[2]{China-Japan Union Hospital, Jilin University, Changchun, 130031, China}
\address[3]{Engineering Research Center of Knowledge-Driven Human-Machine Intelligence, MOE, China}
\address[4]{Department of Computer Science, University of Hong Kong, Pokfulam, Hong Kong}


\begin{abstract}
Spinal curvature estimation is important to the diagnosis and treatment of the scoliosis.
Existing methods face several issues such as the need of expensive annotations on the vertebral landmarks and being sensitive to the image quality.
It is challenging to achieve robust estimation and obtain interpretable results, especially for low-quality images which are blurry and hazy.
In this paper, we propose B-Spine, a novel deep learning pipeline to learn B-spline curve representation of the spine and estimate the Cobb angles for spinal curvature estimation from low-quality X-ray images.
Given a low-quality input, a novel SegRefine network which employs the unpaired image-to-image translation is proposed to generate a high quality spine mask from the initial segmentation result.
Next, a novel mask-based B-spline prediction model is proposed to predict the B-spline curve for the spine centerline.
Finally, the Cobb angles are estimated by a hybrid approach which combines the curve slope analysis and a curve-based regression model.
We conduct quantitative and qualitative comparisons with the representative and SOTA learning-based methods on the public AASCE2019 dataset and our new proposed CJUH-JLU dataset which contains more challenging low-quality images.
The superior performance on both datasets shows our method can achieve both robustness and interpretability for spinal curvature estimation. 
\end{abstract}


\begin{keyword}
\newline Spine curvature estimation 
\newline Computer-aided Cobb angle measurement
\newline B-spline curve
\newline Spine representation
\end{keyword}

\end{frontmatter}
\section{Introduction}
Scoliosis is a condition of spinal deformity which can lead to a side-to-side spine curve.
From an anterior-posterior (AP) view X-ray image, the spine of a person with scoliosis looks like an \textit{S} or a \textit{C} shape rather than a straight line.
Serious scoliosis can affect the breathing and movement of the patients and may even cause the disability if they are not treated in time.
A quantitative measurement of the spinal curvature is important to diagnose the scoliosis and provide the guidance for proper treatment.
Conventionally, the spinal curvature estimation is operated as the Cobb angle measurement in the clinical practice.
Given an AP view X-ray spine image of a person in standing position, the Cobb angles, main thoracic (MT), proximal thoracic (PT) and thoracolumbar/lumbar (TL), can be measured by identifying the most tilted vertebra and computing the angles between the lines drawn from the endplates of involved vertebra.
However, manual measurement of Cobb angles requires to carefully select the end vertebra by a professional clinician, while the process is time-consuming and could lead to subjective errors due to high inter- and intra-observer variability \cite{gstoettner2007inter}. 

In recent years, computer-aided spinal curvature or Cobb angle estimation has been widely studied.
Early methods such as \cite{zhang2010computer,samuvel2012} employ traditional image processing techniques to identify the upper and lower end vertebra for computing the Cobb angles.
These methods mainly have limited accuracy and may be affected by subjective errors.
With the development of machine learning and deep learning, data-driven approaches have achieved promising results by learning Cobb angle estimation based on annotated datasets such as AASCE2019 \cite{AASCE}.
Existing learning-based methods mainly predict the Cobb angles based on certain intermediate representations of the spine including spine segmentation mask, vertebral landmarks and the centerline of the spine.
Despite relatively accurate estimation can be obtained by these learning-based methods, they are still limited in the \textit{robustness} or may lack the \textit{interpretability} of the results.

\begin{figure}
\centering
\includegraphics[width=0.95\linewidth]{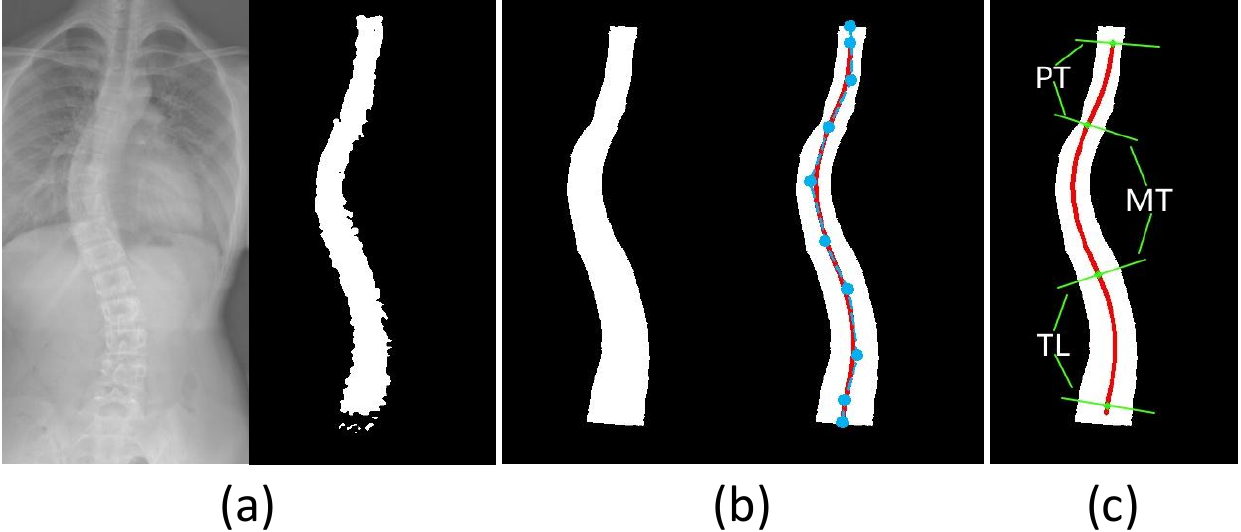}
\vspace{-4pt}
    \caption{
    Given a low-quality X-ray image, our method can robustly learn a B-spline curve representation for the spine centerline and estimate the Cobb angles in an interpretable manner. (a) Input X-ray image and the initial spine mask obtained by an existing segmentation network; (b) Refined mask and the learned B-spline (note the blue points are the predicted control points); (c) Estimated location of endplates for computing the Cobb angles.
    }
  \label{fig:teaser}
\end{figure}

For example, methods like \cite{wang2020scg,lin2021seg4reg+} which directly regress the Cobb angles from the segmentation mask cannot explain where the scoliosis occurs. 
In contrast, the landmarks-based methods \cite{wu2017automatic, lin2020seg4reg}  aim to reconstruct the shape of each vertebra which can lead to well interpretable results.
However, predicting accurate landmarks is challenging since the performance of the model highly relies on the expensive landmarks annotation on the training data and the results are also sensitive to the image quality.
Moreover, for low-quality images which look blurry and hazy as Figure \ref{fig:teaser} (a), it is difficult to annotate the ground truth landmarks of each vertebra even by human.
Hence, the landmarks-based methods may not be suitable or robust for low-quality images.
On the other hand, it is possible to attain both the robustness and interpretability by estimating the centerline of the spine and then predicting the Cobb angles based on the curvature analysis.
Yet, existing centerline-based methods such as \cite{tu2019automatic,dubost2020automated} usually represent the centerline as a fitted polynomial curve or just a collection of connected points, and such curve representations are sensitive to the noise in the input images.
A more proper curve representation which is smooth, stable and easy to learn is needed to improve the robustness of the centerline-based methods.

In this paper, we propose B-Spine, a novel deep learning pipeline to learn B-spline curve representation of the spine centerline and estimate the Cobb angles from low-quality X-ray spine images.
Our goal is to achieve both robustness and interpretability for the spinal curvature estimation by representing the spine with the powerful and well-suited B-spline \cite{unser2002splines}.
Our pipeline consists of three modules: Spine Mask Segmentation, B-Spline Centerline Prediction and Cobb Angle Estimation.
Specifically, we first fine-tune a pretrained segmentation model \cite{shi2022ssformer} to obtain an initial mask of the spine.
Due to the low quality of the input, the initial mask may contain numerous noise which may affect the following centerline learning.
To solve this issue, we enhance the segmentation module with a novel SegRefine network which employs the unpaired image-to-image translation, i.e., CycleGAN \cite{zhu2017unpaired}, to generate a high-quality segmentation mask based on the initial mask.
Next, we learn a B-spline curve representation for the centerline from the segmentation mask based on a novel mask-based B-spline prediction model, by which the control points and knots for constructing the B-spline curve are jointly predicted under supervision from the GT parameters and a new point-based resample loss.
Finally, we propose a hybrid approach which combines the curve slope analysis and a curve-based regression model to obtain the final Cobb angles.

Comparing to the methods based on regressing the Cobb angles directly from the segmentation mask, our centerline-based methods can produce interpretable results which can clearly identifying the place for making the Cobb angle measurement.
Unlike the landmarks-based methods, our B-Spine framework only needs the manual annotation on spine segmentation mask and Cobb angles, while other supervision for the B-spline learning can be robustly derived from the segmentation mask.
Therefore, our method is more applicable and more robust to the challenging low-quality images than the landmarks-based methods.
Similar to other centerline-based methods, we obtain the curve representation of the spine and perform the curve-based Cobb angle estimation.
However, we utilize the B-spline which is a smooth and stable curve representation to represent the spine.
Moreover, we propose a multi-step learning-based B-spline prediction module which is more robust to the image noise and the error from the initial centerline points.

To evaluate the performance of our framework, we perform experiments using the 609 images in the public AASCE2019 dataset and compare the results with representative and SOTA methods which are based on segmentation mask, landmarks and centerline.
In addition, we collect and annotate a more challenging new dataset, named CJUH-JLU (China-Japan Union Hospital, Jilin University) dataset, which contains 584 low-quality X-ray spine images for evaluating existing methods and ours.
Extensive quantitative and qualitative results show our B-Spine can generally outperform the compared methods.
We also conduct ablation studies to verify our proposed modules and design choices, including the SegRefine network, B-spline curve representation and the hybrid scheme for Cobb angle estimation.

In summary, our contribution are as follows:
\begin{itemize}
\item  We propose B-Spine, a novel deep learning pipeline to learn B-spline curve representation of the spine centerline and estimate the Cobb angles from low-quality X-ray images.
To our best knowledge, we are the first to represent the spine using the B-spline curve and our novel multi-stage framework can predict the B-spline and Cobb angles in a robust and interpretable manner. 
\item We propose a novel SegRefine network that employs the unpaired image-to-image translation to generate a high-quality spine mask from the initial segmentation result. With the refined mask, our B-spline prediction is less affected by the quality of the input image.
\item We conduct quantitative and qualitative comparisons with the representative and SOTA learning-based methods on the public AASCE2019 dataset and our new proposed CJUH-JLU dataset.
From the results, our B-Spine framework achieves superior performance especially for challenging low-quality images, demonstrating our method can attain both robustness and interpretability for spinal curvature estimation. 

\end{itemize}

\section{Related Work}
In this section, we first review traditional methods for Cobb angle estimation.
Then, we summarize recent data-driven learning-based methods which incorporate the segmentation mask, vertebral landmarks and spine centerline for spinal curvature estimation.
In the end, we provide a brief survey on the application of B-spline representation in related fields.

\textbf{Traditional methods.}
Most of the early methods are based on the traditional image processing techniques and work in a semi-automatic manner.
In \cite{zhang2010computer,samuvel2012,shaw2012use,akbar2013use,wu2014reliability}, the upper and lower end vertebra of spine are first selected manually and then system can automatically estimate the Cobb angles.
To identify the end vertebra automatically, \cite{prabhu2012automatic} utilizes active contour segmentation and morphological operators to extract the vertebral boundary and computes the Cobb angles based on the slope of the boundary lines.
Similarly, \cite{anitha2014automatic} proposes an automated system to extract the vertebral boundary using a set of customized image filters and estimate the Cobb angles based on the Hough transform.
These non-learning methods can reduce the manual efforts to a certain degree, but they may still suffer from subjective errors and are not robust to noisy inputs.

\textbf{Estimation based on the spine segmentation mask.}
Thanks to the rapid development of segmentation techniques \cite{ronneberger2015u,huang2020unet}, the spine segmentation mask can be extracted in a relatively accurate and efficient manner.
Based on the learned segmentation mask, \cite{wang2020scg} trains an angle regression network to predict the Cobb angles.
Instead of training the segmentation and angle regression networks separately, Seg4Reg+ \cite{lin2021seg4reg+} jointly optimizes these two networks and achieves the SOTA performance for Cobb angle estimation.
As these methods mainly perform the prediction by direct regression on the segmentation mask, their results are not well interpretable, i.e., it is unknown which image region the angles are corresponding to.
In our B-Spine, the segmentation mask is used to predict the B-spline centerline which can be processed to obtain interpretable angle estimation.
Furthermore, to improve the noisy segmentation mask obtained from the low-quality X-ray image, we take the inspiration from \cite{gu2021cyclegan} and employ the unpaired image-to-image translation model CycleGAN \cite{zhu2017unpaired} to generate refined mask with most of noise removed.
With the refined mask, we can perform more robust mask-based B-spline prediction. 

\textbf{Estimation based on vertebral landmarks.}
Besides the spine segmentation mask, the shape of each vertebra can provide more geometric and structural information of the spine.
Normally, the four corners of each vertebra are annotated as the landmarks which can be predicted from an input X-ray image and be used to estimate the Cobb angles.
The landmarks prediction is usually treated as a regression task.
In \cite{sun2017direct}, the structured support vector regression ($\mathrm{S^2VR}$) is employed to jointly estimate the landmarks and Cobb angles in a single framework using the traditional HOG feature descriptor.
\cite{wu2017automatic} proposes the BoostNet which integrates statistical outlier removal methods into CNN-based feature extraction to obtain robust image features for landmarks regression.
The data from \cite{wu2017automatic} which contain landmarks annotations on 17 vertebra composed of the thoracic and lumbar spine later become the base of the AASCE2019 dataset \cite{AASCE} and promote the advancement of landmarks-based spinal curvature estimation \cite{chen2019automated,lin2020seg4reg,kim2020automation,zhang2021automated}.
In addition to predicting the landmarks from the single AP view X-ray images, \cite{wu2018automated,wang2019accurate} utilize multi-view X-ray images for more robust Cobb angle estimation.
Due to the annotation difficulty and estimation error, it is hard to obtain accurate landmarks for low-quality images.
In contrast, we leverage the segmentation mask which is easier to annotate and extract, and conduct mask-based B-spline centerline prediction. 


\textbf{Estimation based on the spine centerline.}
Since the spine itself is in a curve shape, the spinal curvature estimation can be directly performed on the spine centerline curve extracted based on input X-ray image, segmentation mask or the extracted vertebra shape.
In general, the centerline can be represented in different curve representations.
For example, \cite{alharbi2020deep} directly connects the center points of the vertebra estimated from the input X-ray image using CNN.
In \cite{tu2019automatic}, the centerline curve is obtained by fitting a 6-degree polynomial curve to the center points extracted based on the spine segmentation mask.
To improve the robustness of center points extraction, \cite{dubost2020automated} proposes CasNet which employs cascaded segmentation networks to obtain the segmentation mask of the centerline, from which the dense center points are extracted, connected and smoothed to form the centerline curve.
For above methods, the Cobb angles can be obtained by analyzing the slopes of centerline curve.
On the other hand, \cite{huo2021joint} jointly optimizes the centerline extraction and the centerline-based Cobb angle regression.
Comparing to above methods, we adopt the B-spline curve for more robust centerline extraction and design a specific network to learn the B-spline parameters from the segmentation mask.
To further improve the robustness and interpretability of our pipeline, we utilize a hybrid approach to combine the results from the curve slope analysis and curve-based Cobb angle regression.

\textbf{B-spline representation and its application.}
B-spline is a powerful curve or surface representation which is widely used in computer graphics and computer-aided design \cite{gordon1974,patrikalakis2002shape}.
In the medical imaging domain, B-spline representation has also been investigated for image interpolation \cite{lehmann1999survey}, segmentation \cite{brigger2000b} and registration \cite{klein2007evaluation} etc.
By using B-spline to represent the curves, the results are usually smoother and more robust \cite{williams2021interactive}. 
Meanwhile, to our best knowledge, there is no prior work that uses the B-spline to represent the spine centerline.
Recently, some works apply deep learning to predict B-spline representation from point sequence \cite{laube2018deep} and segmentation boundary \cite{barrowclough2021binary}.
We take the similar idea of using deep learning to predict the B-spline parameters, while our network is designed for the mask-based centerline approximation.

\begin{figure*}[t]
\centering
\includegraphics[width=0.95\textwidth]{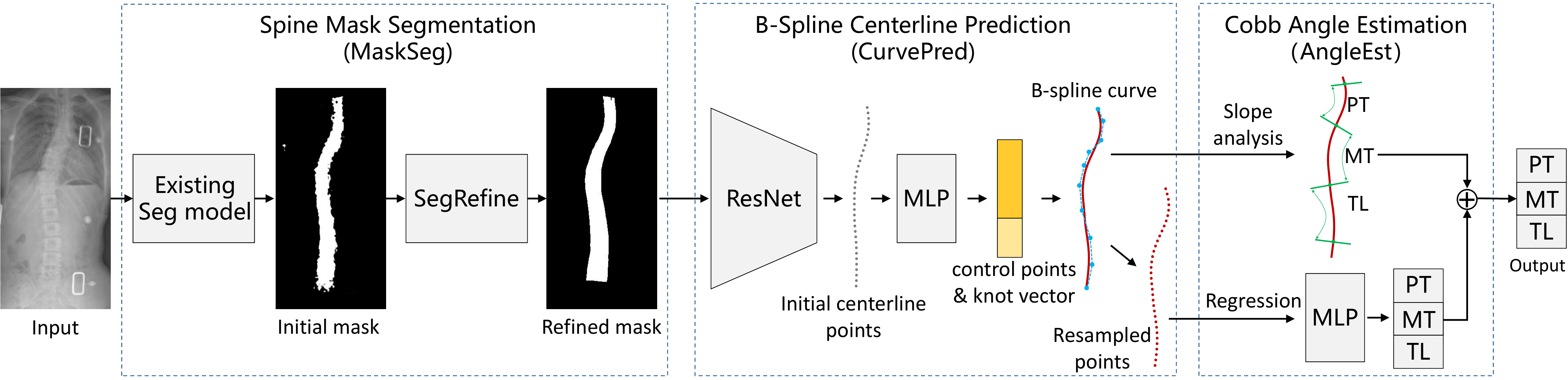}
\vspace{-4pt}
\caption{
Overview of our B-Spine pipeline for estimating Cobb angles from a low-quality X-ray image.
}
\vspace{-4pt}
\label{fig.pipeline}
\end{figure*}

\section{Method}

In this section, we first briefly review the preliminaries on B-spline curve including the mathematical definition and construction algorithm.
Then, we introduce details about the three modules of our B-Spine framework.
Figure \ref{fig.pipeline} shows an overview of our multi-stage pipeline.

\subsection{Preliminaries on B-Spline Curve}
Spline curve is a smooth curve widely used for curve fitting or interpolation.
It is defined by using piecewise polynomials to approximate or pass through a series of given points. 
B-spline or basis spline curve is a specific type of spline curve which has good properties such as local support, flexibility and high approximation accuracy.
Given n + 1 control points $\{P_0, P_1, \cdot\cdot\cdot, P_n\}$ and a knot vector $\mathbf{U} = \{ u_0, u_1, \cdot\cdot\cdot, u_m \}$, the B-spline curve of degree $p$ is defined as:
\begin{equation}
    C(u)=\sum_{i=0}^n B_{i,p}(u) P_i,
    \label{form.bspline}
\end{equation}
where $B_{i,p}(u)$ are B-spline basis functions of degree $p$. 
The most common way to define the B-spline basis functions is using the Cox-de Boor recursion formula:

\begin{equation}
    B_{i,0}(u)=
    \begin{cases}
        1, &  \text{if} \; u_i \leq u \lq u_{i+1}, \\
        0, &  \text{otherwise}.
    \end{cases} 
    \label{form.deboor_0}
\end{equation}

\begin{equation}
    \resizebox{.9\hsize}{!}{
    $B_{i,p}(u)=\frac{u-u_i}{u_{i+p}-u_i} B_{i,p-1}(u) +
        \frac{u_{i+p+1}-u}{u_{i+p+1}-u_{i+1}} B_{i+1,p-1}(u).$
    }
    \label{form.deboor_p}
\end{equation}
Based on the definition of B-spline, the following relationship between the degree $p$, the number of control points $n+1$ and the length of knot vector $m+1$ needs to be satisfied: $m=n+p+1$.
In this paper, we adopt a special case of B-spline, i.e., clamped B-spline , for which the curve passes through the first and last control points and is also tangent to the first and last legs of its control polyline.
To enable the clamped effect, it is required the first $p+1$ and the last $p+1$ knots must be identical.
Please refer to \cite{patrikalakis2002shape} for more introduction about B-spline.



\subsection{Spine Mask Segmentation \label{sec.segmentation} }
The first module of our B-Spine framework is Spine Mask Segmentation or MaskSeg in short.
Given an input X-ray image, we first segment the spine mask with a conventional semantic segmentation network.
Here, we employ SSformer \cite{shi2022ssformer} which is a lightweight transformer-based method and enhance it with an additional structural similarity index (SSIM) loss \cite{zhou2004image} for better segmentation quality.
Other semantic segmentation methods such as U-Net \cite{ronneberger2015u} can also be used in this step to obtain an initial segmentation mask.
However, due to the low quality of X-ray images, there may be significant noise in the initial mask (see Figure \ref{fig:teaser} (a)).
Also, according to our experiments, it is difficult to reduce the noise by simply fine-tuning the network or using other segmentation models.

Inspired by the application of generative adversarial networks (GANs) for denoising low quality medical images \cite{gu2021cyclegan}, we propose a SegRefine network which refines the initial noisy mask based on the unpaired image-to-image translation technique.
Specifically, we train a CycleGAN model \cite{zhu2017unpaired} with a set of noisy masks predicted by the conventional segmentation network and a set of GT masks, while the images of these two sets are unpaired which can improve the generalization ability of the SegRefine network.
After training, the SegRefine network can translate/generate a refined mask from the initial noisy mask, or vise versa.
Comparing to the supervised learning based method, our unsupervised SegRefine model will mainly focus on converting images in one set into the \textit{style} of images in the other set.
Thus, it is more robust and more generalizable to the more diversified masks obtained from low quality images.

\subsection{B-Spline Centerline Prediction\label{sec.bspline}}
The second module of our pipeline is B-spline Centerline Prediction or CurvePred, which predicts the B-spline parameters for representing the spine centerline from the segmentation mask.
In our implementation, we choose the clamped B-spline of degree $p=3$ and the number of control points and knots are 10 and 14, respectively.
Following the constraints for the clamped B-spline, the first 4 knots are set to be 0 and the last 4 knots are set to be 1, i.e., the knot vector become to $\{0,0,0,0,u_4,u_5,\cdot\cdot\cdot,u_9,1,1,1,1\}$.
Hence, the B-spline parameters to predict include positions of 10 control points and values of 6 knots.
We formulate the B-spline prediction as a regression problem and use GT values as supervision to compute losses for these parameters.
Besides, we also propose a new \textit{resample loss} which is computed based on the centerline points resampled from the predicted and GT B-spline to enforce explicit point supervision on the B-spline curve. 
In Section \ref{sec.datasets}, we provide details for how to prepare the GT centerline points.
Overall, the CurvePred module contains three steps: 1) predict initial centerline points from the refined segmentation mask; 2) predict the B-spline parameters; 3) resample centerline points from the predicted B-spline and use them to compute the resample loss.

\textbf{Initial centerline points prediction.}
Directly predicting the B-spline parameters from the segmentation mask may lead to unstable results.
Therefore, we first predict a set of initial centerline points from the mask and then use these points as input for predicting the B-spline parameters.
To predict the centerline points, we modify the last layer of ResNet101 \cite{he2016deep} so that the coordinates of 34 centerline points are predicted under the supervision of GT centerline points.
The loss for predicting initial centerline points is defined as:
\begin{equation}
L_{init}=\frac{1}{34}\sum_{i=1}^{34} \|C_{i}^{init}- C_{i}^{GT}\|^2_2,
\label{form.init}
\end{equation}
where $C_{i}^{init}$ and $C_{i}^{GT}$ denote i-th predicted or GT centerline points and $\|\cdot\|_2$ is the L2 norm.

\textbf{B-spline parameters prediction.}
From initial centerline points, we use a 4-layer MLP to predict the B-spline parameters, i.e., the control points and knots, and define the loss:
\begin{equation}
    \resizebox{.9\hsize}{!}{
$L_{paras}=\frac{1}{10}\sum_{i=0}^{9} \|P_{i}^{pred}- P_{i}^{GT}\|_2^2
 +\frac{1}{6}\sum_{i=4}^{9} \|U_{i}^{pred}- U_{i}^{GT}\|_2^2,$
 }
\label{form.paras}
\end{equation}
where $P_{i}^{pred}$, $P_{i}^{GT}$, $U_{i}^{pred}$, $U_{i}^{GT}$ denote the i-th predicted or GT control points or knots, respectively.


\textbf{Resample loss.}
After predicting the control points and knots, we construct the B-spline curve $C(u)$ using the Cox-de Boor algorithm.
As the B-spine parameters are predicted from the initial centerline points, the accuracy of the curve may be affected by the quality of the segmentation mask.
To improve the robustness of curve prediction and boost the performance, we propose a new resample loss to introduce more constraints on the centerline points computed from the predicted curve. 
Specifically, we use the same set of parameters $\{u_i\}$ for generating the GT centerline points and obtain a set of resampled centerline points $\{C(u_i)\}$.
Then, the resample loss is defined as:
\begin{equation}
L_{resample}=\frac{1}{34}\sum_{i=1}^{34} \|C(u_i)- C_{i}^{GT}\|_2^2, \\
\label{form.l_{resample}}
\end{equation}
where $u_i$ is the i-th curve parameter obtained by uniform sampling from $[0,1]$, following the same procedure as for GT.


In summary, the total loss for CurvePred module is:
\begin{equation}
\label{form.loss}
    L_{CurvePred} = \lambda_1 \cdot L_{init} + \lambda_2 \cdot L_{paras} + \lambda_3 \cdot L_{resample}.
\end{equation}
In our current implementation, we set $\lambda_1=1$, $\lambda_2=0.1$, $\lambda_3=0.1$ and jointly train the ResNet101 and the MLP for initial centerline points and B-spline parameter prediction.

\subsection{Cobb Angle Estimation\label{sec.cobb}}
The last module of our pipeline is Cobb Angle Estimation or AngleEst.
From the predicted B-spline centerline, the Cobb angles can be directly estimated by performing curve slope analysis.
On the other hand, as a multi-stage pipeline, the errors from previous modules may accumulate and lead to inaccurate B-spline curve.
To alleviate this issue, we further learn a regression model to predict the Cobb angles from the learned B-spline curve.
Then, we combine the results from both schemes and obtain the final Cobb angle estimation.

\textbf{Cobb angles from curve slope analysis.}
Similar to previous methods \cite{dubost2020automated}, we can estimate the Cobb angles by finding the maximum and minimum slopes of the spine curve at sampled points.
Given a predicted B-spline curve, we first sample 17 points, denoted as $\{S_i\}$, with equal arc length, while these points correspond to 12 thoracic vertebra and 5 lumbar vertebra.
Next, for each point $S_i$, we sample another point $S'_i$ that is sufficiently close to it so that we can compute the slope at $S_i$ as:
\begin{equation}
slope_i=\frac{S'_{i,y}-S_{i,y}}{S'_{i,x}-S_{i,x}},
\label{form.slope}
\end{equation}
where $S_{i,x}$, $S'_{i,x}$, $S_{i,y}$, $S'_{i,y}$ are the x and y coordinates for $S_i$ and $S'_i$, respectively.
Here, we assume the spine spans along the y direction and all $S'_i$ are sampled at the same neighboring side of $S_i$.
Then, we find the maximum and minimum slopes w.r.t the 17 sampled points and compute the MT angle as:
\begin{equation}
Cobb^s_{MT}=\frac{180}{\pi}|\arctan \frac{slope_{max}-slope_{min}}{1+slope_{max} \cdot slope_{min}}|
\label{form.cobb}
\end{equation}
Similarly, we can compute the PT or TL angle by finding the maximum or minimum slopes above or below the points used for computing MT.
The results of the slope-based Cobb angles are denoted as $Cobb^s=\{Cobb^s_{MT},Cobb^s_{PT},Cobb^s_{TL}\}$.

\textbf{Cobb angle regression.}
To regress the Cobb angles from the B-spline centerline, we reuse the 34 resampled points obtained from Section \ref{sec.bspline} as input and train a 4-layer MLP to predict the angles under the supervision of GT values, while MSE loss is used for each angle.
The results of the predicted Cobb angles are denoted as $Cobb^r=\{Cobb^r_{MT},Cobb^r_{PT},Cobb^r_{TL}\}$.

Through our statistical analysis, $Cobb^s$ and $Cobb^r$ have different error distributions. When one method produces an over-estimated result w.r.t. to GT angle, the other one may produce an under-estimated result. 
Finally, we linearly combine the slope-based and regression-based Cobb angles and obtain the final results:
\begin{equation}
Cobb=\alpha \cdot Cobb^s+(1-\alpha) \cdot Cobb^r,
\label{form.final_cobb}
\end{equation}
where $\alpha$ can be selected based on the performance of each angle using the validation set of different datasets.
From our experiments, we set $\alpha=0.4$ for MT, and $\alpha=0.5$ for PT and TL on AASCE2019 dataset, and  $\alpha=0.5$ for MT, and $\alpha=0.8$ for PT and TL on CJUH-JLU dataset.
More details about the choice of parameter $\alpha$ are provided in Section \ref{result:ablation}.

\section{Results}
In this section, we first provide the details of two datasets and evaluation metrics we used for experiments.
Then, we perform quantitative and qualitative comparisons with representative and SOTA methods to show the superiority of our method.
Ablation studies are also conducted to verify the effectiveness of proposed modules.

\subsection{Datasets and Evaluation Metrics}
\label{sec.datasets}

\textbf{AASCE2019 dataset.}
The MICCAI 2019 Challenge on Accurate Automated Spinal Curvature Estimation provides the dataset which is formed based on data from \cite{wu2017automatic,wang2019accurate}.
The dataset consists of 609 spinal AP X-ray images with 481 images used for training and 128 images for testing.
Each image is annotated with the corners of each vertebra as landmarks (17$\times$4 in total) and three Cobb angles (MT, PT and TL) by professional clinicians.
To utilize this dataset in our framework, we first obtain the spine segmentation mask by connecting the left and right boarders of each vertebra to get the left and right mask contours, while the top and bottom boarders of the first and last vertebra become the top and bottom contours.
Then, we extract the center of top and bottom boarders of each vertebra and get 34 initial centerline points.
Then the GT parameters for the B-spline, including 10 control points, 14 knots (with 6 unknown knots), can be estimated by fitting the 3-degree clamped B-spline to the initial centerline points.
In addition, we sample 34 GT centerline points from the B-spline curve by uniform sampling the curve parameter $u \in [0,1]$.

\textbf{CJUH-JLU dataset.}
To evaluate the performance of our method in more challenging scenarios, we build a new dataset which contains 584 spinal AP X-ray images collected from China-Japan Union Hospital, Jilin University.
We split the dataset by 7:1:2, i.e., 408, 58 and 118 images for training, validation and testing, respectively.
These images are normally in a lower quality than AASCE2019, e.g., blurrier and hazier, and it may be difficult to identify individual vertebra even by human.
Thus, it is almost impossible to annotate the vertebral landmarks and train the landmarks-based methods on this dataset.
On the other hand, it is easier to annotate the segmentation mask since only the rough shape needs to be labelled and the professionals have sufficient knowledge to identify the spine even from blurry images.
Therefore, we design a labelling tool based on LabelMe \cite{russell2008labelme} for manual annotation.
Two professional local clinicians are trained to use our tool for annotating each image in CJUH-JLU with the spine segmentation mask and three GT Cobb angles.

\textbf{Evaluation metrics.}
Following previous works \cite{lin2020seg4reg,dubost2020automated}, we evaluate the accuracy of the results using the mean absolute error (MAE) for each Cobb angle, in degrees. In addition, the symmetric mean absolute percent error (SMAPE) for all Cobb angles are computed:
\begin{equation}
SMAPE=\frac{1}{N}\sum_{i=1}^{N}\frac{\sum_{\theta \in \Theta}^3|A_{i,\theta }-A'_{i,\theta }|}{\sum_{\theta \in \Theta}^3{(A_{i,\theta }+A'_{i,\theta})}} \times 100 \%,
\label{form.smape}
\end{equation}
Here, $\Theta=\{MT, PT, TL\}$ is the set of Cobb angles; $A_{i,\theta}$ and $A'_{i,\theta}$ are the ground truth and predicted Cobb angles, respectively; $N$ is the number of images in the evaluation set.

\subsection{Implementation Details}
\label{result:implementation}
\begin{figure}[t]
    \centering
    \includegraphics[width=0.95\linewidth]{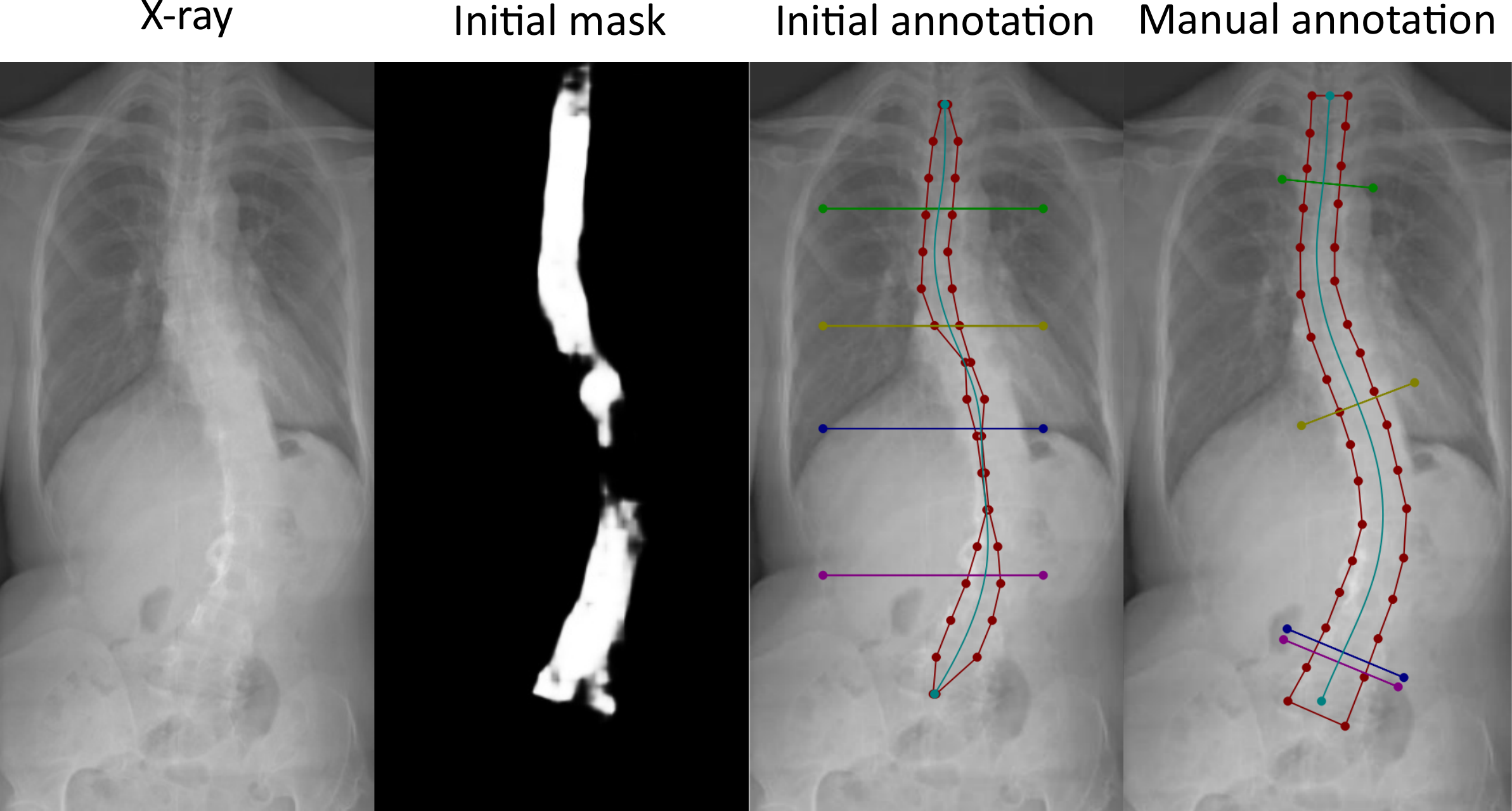}
    \caption{
    Illustration of the data annotation process (see Section \ref{result:implementation} for details about the process).
    }
    \label{fig:annotation}
\end{figure}

\textbf{Data annotation details.}
Figure \ref{fig:annotation} shows an illustration of our data annotation process.
Given an X-ray image cropped and/or resized to 512$\times$256 resolution, we first use the re-trained SSformer \cite{shi2022ssformer} to obtain the initial segmentation mask which may be noisy due to the low-quality input.
Then, an initial mask annotation with 34 contour points is provided based on the initial segmentation mask.
Each of the left and right contours contains 17 points and a user can drag the individual point to adjust the shape of the mask.
Meanwhile, we compute 17 initial centerline points by evenly sampling from the midpoints of the segments generated by performing horizontal pixel-wise scanning along the left and right mask contours.
These initial centerline points are used to fit a clamped B-spline with 10 control points and 14 knots (with 6 unknowns).
Such control points and knots are used as the GT for learning the B-spline parameters.
In addition, we sample 34 GT centerline points from the fitted B-spline curve by uniformly sampling the curve parameter $u$.
To further assist the user in mask annotation, we visualize the centerline with connected GT centerline points in real-time when the user is adjusting the mask contour points.
Such visual hints can help the user obtain both more plausible segmentation mask and the fitted B-spline centerline.
Besides the segmentation mask, our tool also allows users to annotate the GT Cobb angles by adjusting the position and orientation of the handles representing the key vertebral endplates.

\textbf{Training details.} Our framework contains several networks which are trained in different manners. 
In MaskSeg, input X-ray images are resized to 512$\times$256 following \cite{lin2020seg4reg} and the SSformer is trained using the GT masks as supervision to produce the initial segmentation masks.
Then, we train the CycleGAN using the initial segmentation masks and the GT masks as two sets of unpaired data, and the learned generator for the initial-to-refine mask translation is used as the SegRefine network.
In CurvePred, we train the ResNet101 based image encoder and the MLP jointly, while the GT centerline points and GT B-spline parameters are used as supervision.
In AngleEst, the MLP for regressing the Cobb angles is trained using the resampled B-spline centerline points as input and the GT angles as supervision.
Note all networks are trained using the images in the currently evaluated dataset.

When training our pipeline, data augmentation including random flipping, rotation (between -10$^\circ$ and 10$^\circ$) and re-scaling (with factor between 0.85 and 1.25) is applied to the input images, masks, as well as the centerline points and B-spline parameters to train the spine mask segmentation, B-spline prediction and Cobb angle regression modules. 
Our code is implemented in PyTorch and runs on a PC with a 3.6GHz CPU and one NVIDIA RTX 3060 GPU.
For all the networks, we use the Adam as the optimizer, and the training setting and time for each module are shown in Table \ref{tab:training-setting}.
It can be seen the B-spline learning takes longer time than other modules.
The main reason is by considering the resample loss, we need to perform points resampling after the B-spline is updated in each learning step.
How to impose the point-wise supervision while accelerating the B-spline learning process is an interesting future work.

\begin{figure}[t]
    \centering
    \includegraphics[width=0.98\linewidth]{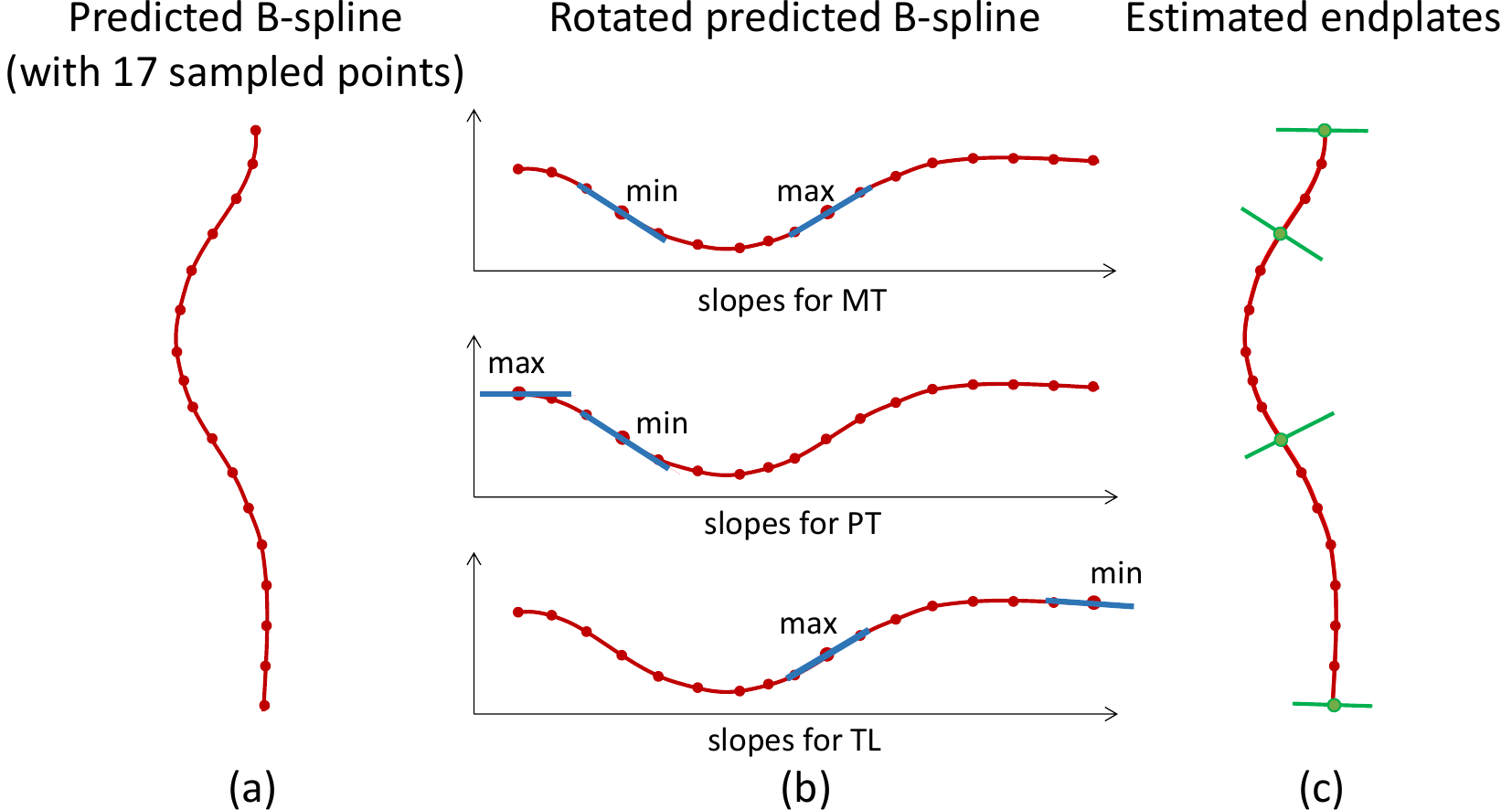}
    \caption{
    Illustration of using the curve slope analysis for Cobb angle estimation.
    The predicted B-spline is first rotated by 90 degrees anticlockwise. 
    Then, slopes at 17 uniformly sampled points are computed.
    The maximum and minimum slopes of the whole curve are used to compute MT.
    The maximum and minimum slopes above or to the left of (from the rotated view) the previous points are used to compute PT.
    TL is computed in a similar way as PT.
    }
    \label{fig:slope}
\end{figure}

\textbf{Inference details.} During the inference, we perform curve slope analysis and angle regression to estimate the Cobb angles.
For curve slope analysis, we sample 17 points $\{S_i\}$ with parameters $\{u_i\}$ and another 17 points $\{S_i'\}$ with $\{u_i+\epsilon \}$ from the predicted B-spline curve, while the hyperparameter $\epsilon$ is set to a small value 5e-2.
Before computing the slopes of the line $[u_i, u_i+\epsilon]$, we rotate the curve by 90 degrees anticlockwise so that the slopes do not contain infinity and negative infinity values.
Then, the slopes and Cobb angles are computed following Equation \ref{form.slope} and \ref{form.cobb}, as well as the steps described in Section \ref{sec.cobb}.
In Figure \ref{fig:slope}, we show an illustration of using the curve slope analysis for the Cobb angle estimation.


\begin{table}[!t]
    \caption{Training details for different modules of our pipeline. For the modules, InitSeg represents the SSformer, SegRefine represents the CycleGAN, B-spline includes the ResNet for obtaining the initial centline points and MLP for predicting the B-spline parameters, CobbReg is the MLP for regressing the Cobb angels from the resampled centerline points.}
    \centering
    \setlength{\tabcolsep}{4pt}
    \begin{tabular}{lcccc}
        \hline
       Module & InitSeg & SegRefine & B-spline & CobbReg \\
        \hline
        Augmentation & \multicolumn{4}{c}{random flip + rotate + resize} \\
        \hline
        LR & 1e-3 & 1e-3 & 1e-4 & 1e-4\\
        LR decay & 0.9, 1000 & - & 0.9, 1000 & 0.9,1000\\
        Batch size & 4 & 2 & 16 & 1024\\
        Epochs & 500 & 200 & 1000 & 1000\\
        Training time & 1h & 5h & 10h & 0.5h\\
        \hline
    \end{tabular}
    \label{tab:training-setting}
\end{table}

\begin{table}
    \caption{Quantitative comparisons on AASCE2019 (upper part) and CJUH-JLU dataset (lower part)}
    \centering
    \setlength{\tabcolsep}{1pt}
    \begin{tabular}{lcccc}
        \hline
        Method & MAE$_{MT}$ & MAE$_{PT}$ & MAE$_{TL}$ & SMAPE \\
        \hline
        Seg4Reg \cite{lin2020seg4reg} & 4.71 & 5.79 & 5.65 & 10.85  \\
        Seg4Reg+ \cite{lin2021seg4reg+} & 3.88 & 4.62 & 4.99  & \textbf{8.47} 
        
        \\
        CasNet \cite{dubost2020automated}  & 5.12 & 5.41 & 5.43 & 11.42  \\
        JointNet \cite{huo2021joint} & 4.26 & 4.87 & 5.18 & 10.25 \\     
        Ours & \textbf{3.73} & \textbf{4.13} & \textbf{4.81} & 9.53 \\
        \hline
        Seg4Reg \cite{lin2020seg4reg} & 4.89 & 5.87 & 6.34 & 18.87  \\
        CasNet \cite{dubost2020automated} & 4.73 & 7.79 & 7.28 & 22.22 \\
        Ours & \textbf{4.17} & \textbf{4.59} & \textbf{4.20} & \textbf{15.55}
        \\
        \hline
    \end{tabular}

    \label{tab.sota.public}
\end{table}

\begin{figure*}[!t]
    \centering 
    \includegraphics[width=0.95\linewidth]{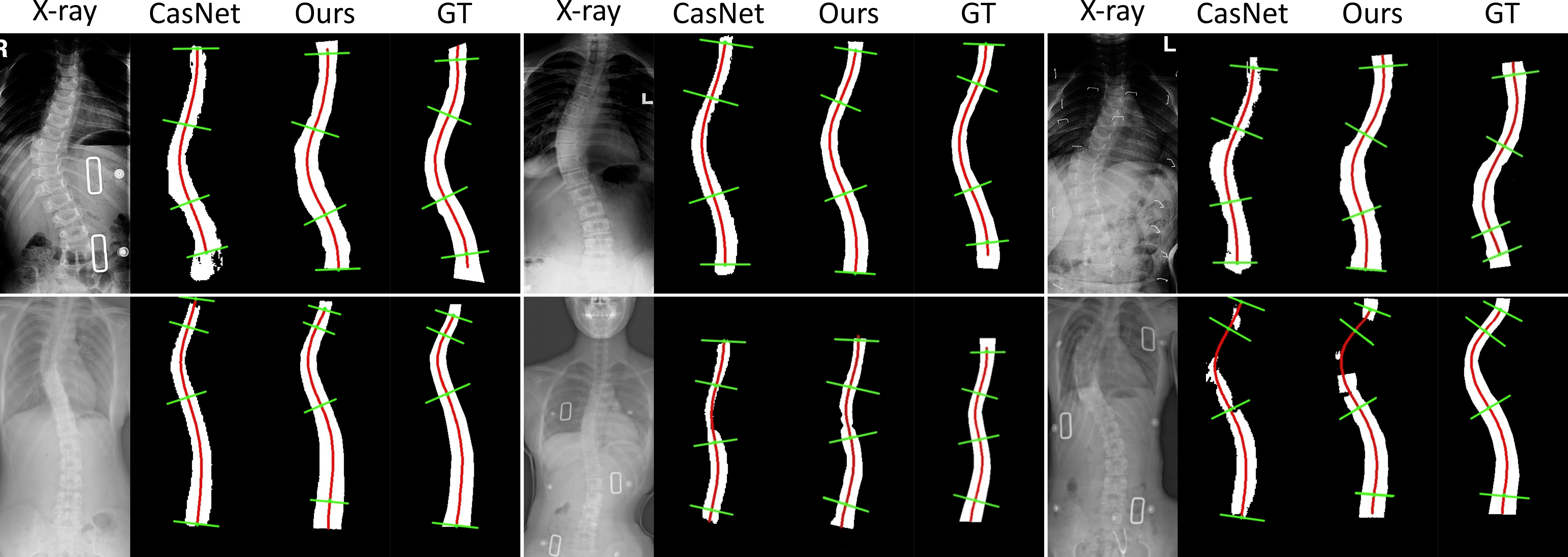}
    \caption{
    Qualitative comparisons on images from AASCE2019 (top) and CJUH-JLU (bottom). 
     It can be observed our centerlines (red curve) and location of endplates (green segments) for estimating the Cobb angles are closer to the GT.
    } 
\label{fig.centerline}
\end{figure*}

\subsection{Comparisons}
\textbf{Quantitative comparisons.}
We perform quantitative evaluations on both AASCE2019 and CJUH-JLU dataset to compare the performance of our B-Spine framework with existing representative and SOTA methods.
For AASCE2019, the compared methods include:
Seg4Reg \cite{lin2020seg4reg}, a representative landmarks-based method which wins the MICCAI 2019 AASCE Challenge;
Seg4Reg+ \cite{lin2021seg4reg+}, a mask-based method which achieves the SOTA performance on AASCE2019;
CasNet \cite{dubost2020automated}, a representative centerline-based method that predicts the dense centerline points based on cascaded segmentation networks;
JointNet \cite{huo2021joint}, a joint centerline prediction and Cobb angle estimation method which achieves the SOTA performance on AASCE2019 among the centerline-based works.
We compare the MAE and SMAPE values reported in their paper.
For CJUH-JLU dataset, we only compare with Seg4Reg and CasNet as only the code for Seg4Reg is available and the CasNet is relatively straightforward to implement.
For Seg4Reg+ and JointNet, we do not try to implement their pipelines as they are quite involved and our implementation may not well reflect the capability of their methods.
To produce the results, we use the released code of Seg4Reg and CasNet implemented by ourselves to train the models with images in CJUH-JLU dataset.

From Table \ref{tab.sota.public}, our method achieves the best performance except the SMAPE on AASCE2019.
One possible reason is the SMAPE metric itself prefers the predicted angles which are larger than GT, i.e., since the predicted values also appear in the denominator (see Equation \ref{form.smape}), SMAPE would produce smaller values when the predicted angle is larger, even if the relative absolute error from a smaller predicted angle is the same to the error of the larger prediction.

\textbf{Qualitative comparisons.}
To quantitatively evaluate our method, in Figure \ref{fig.centerline}, we show the visualization of our results and compare them with the results from the centerline-based method CasNet.
The reason for choosing CasNet for comparison is other methods such as Seg4Reg and Seg4Reg+ are mainly regression-based and non-interpretable and no qualitative results can be visualized.
From the visualized results, our method is able to perform robust and accurate prediction for low-quality images with noisy or  
partially broken masks.

\begin{table}[t]
    \caption{Quantitative ablation studies on AASCE2019 dataset}
    \centering
    \setlength{\tabcolsep}{1pt}
    \begin{tabular}{lccccc}
        \hline
        Method  &MAE$_{avg}$  & MAE$_{MT}$ & MAE$_{PT}$ & MAE$_{TL}$ & SMAPE \\ 
        \hline
        w/o SegRefine    & 4.64     & 4.00    & 4.64     & 5.28     & 10.43  \\ 
        \hline
        6-degree Polynomial & 5.36 & 3.88 & 5.91 & 6.30 & 11.02 \\ 
        CubicSpline & 4.65 & 4.38 & 4.26 & 5.31 & 9.88 \\ 
        CubicHermiteSpline & 4.35 & 4.03 & 4.14 & 4.89 & 9.69 \\ 
        \hline
        $L_{paras}$ only & 5.45 & 4.43 & 5.57 & 6.34 & 12.87 \\
        w/o $L_{resample}$ & 4.75 & 4.13 & 4.87 & 5.26 & 10.53 \\ 
        \hline
        $Cobb^s$ only & 4.73 & 4.37 & 4.66 & 5.16 & 10.89 \\ 
        $Cobb^r$ only & 4.95 & 4.17    & 5.13     & 5.54     & 10.86 \\ 
        \hline
        Ours & \textbf{4.22} & \textbf{3.73} & \textbf{4.13} & \textbf{4.81} & \textbf{9.53} \\ 
        \hline
    \end{tabular}
    \label{tab.ablation}
\end{table}

\subsection{Ablation Studies}
\label{result:ablation}

\textbf{Effectiveness of the SegRefine network.}
The SegRefine network improves the quality of the segmentation mask so that the following CurvePred module receives more refined input.
To verify its effectiveness, we compare with the pipeline trained with the SegRefine network disabled, i.e., the initial segmentation mask from the SSformer is used for predicting the B-spline.
From the comparison between the 2nd and last row of Table \ref{tab.ablation} and \ref{tab.ablation-sanyuan}, it shows the SegRefine network can indeed lead to more accurate Cobb angle estimation.
Furthermore, we quantitatively compare the segmentation masks obtained with different methods.
We evaluate the intersection over union (IoU or Jaccard index), Dice coefficient (Dice), pixel-wise accuracy (pixel-AC) metrics of the output segmentation masks for the AASCE2019 dataset.
The results reported from Seg4Reg+ \cite{lin2021seg4reg+} and our method without the SegRefine network are compared in Table \ref{tab.segmentation}.
It can be seen our SegRefine network can indeed improve the accuracy of the segmentation mask and generate the results outperforming other baselines.

\begin{figure}[t]
    \centering
    \includegraphics[width=0.95\linewidth]{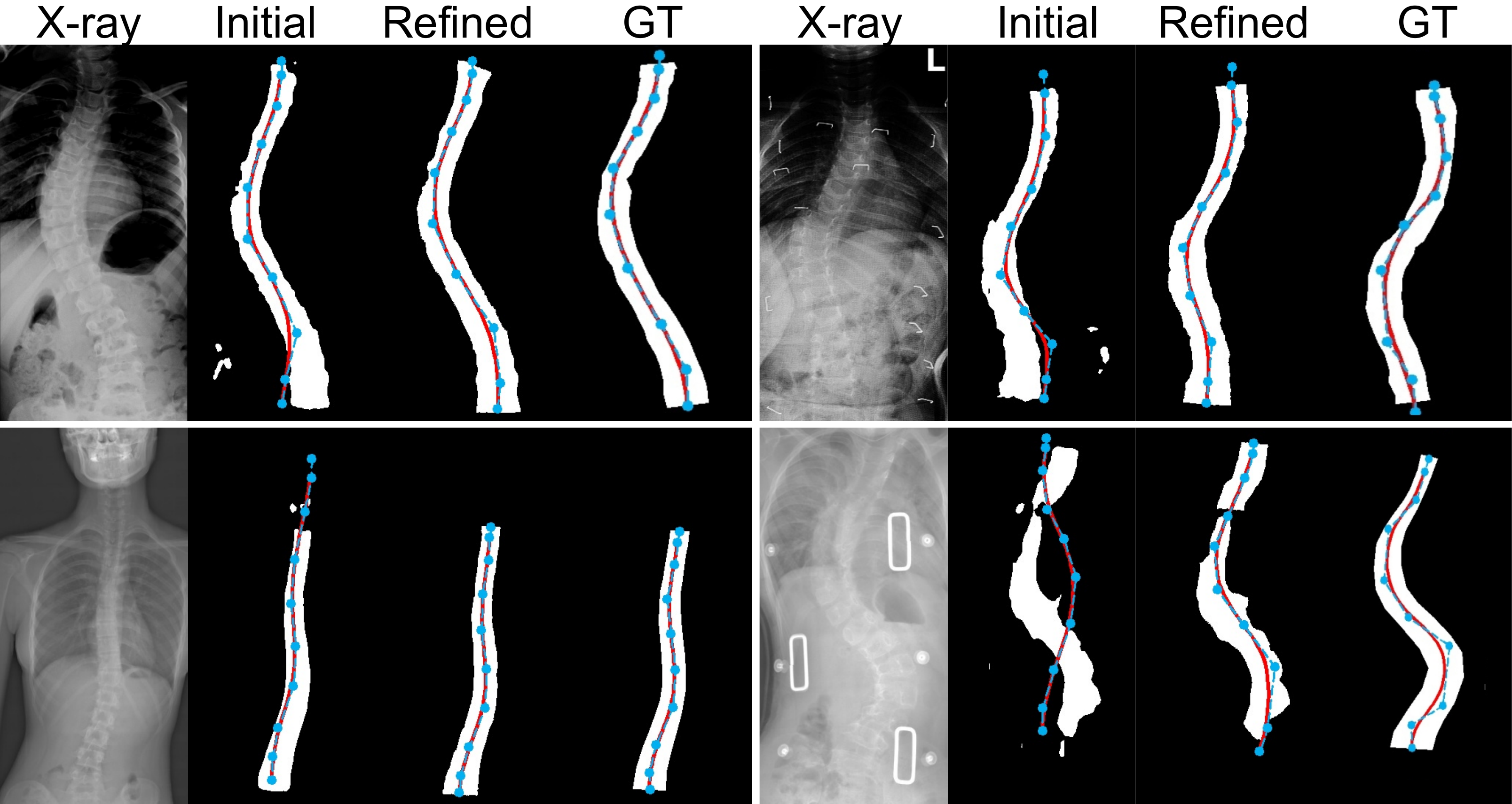}
    \caption{
    Qualitative comparison of the masks and B-spline curves obtained from without and with SegRefine network.
    Images in the top and bottom row are from AASCE2019 and CJUH-JLU datasets, respectively.
    With SegRefine, the masks contain less noise and the B-spline curves (in red) are more closer to GT.
    }
    \label{fig:segrefine}
\end{figure}

We also qualitatively evaluate the effectiveness of the SegRefine module by comparing the predicted B-spline curve obtained from the initial and the refined segmentation masks.
From Figure \ref{fig:segrefine}, it can be seen the initial mask contain noise and outliers which will negatively affect the B-spline prediction.
With the SegRefine network, not only the quality of the mask is improved, the predicted B-spline is also more plausible and closer to the GT.

\begin{table}[t]
    \caption{Quantitative ablation studies on CJUH-JLU dataset}
    \centering
    \setlength{\tabcolsep}{1pt}
    \begin{tabular}{lccccc}
        \hline
        Method   & MAE$_{avg}$ & MAE$_{MT}$ & MAE$_{PT}$ & MAE$_{TL}$ & SMAPE \\
        \hline
        w/o SegRefine & 4.39 & 4.37    & \textbf{4.55}     & 4.24     & 15.86  \\ 
        \hline
        6-degree Polynomial & 5.11 & 4.36 & 6.29 & 4.68 & 16.23 \\ 
        CubicSpline & 4.44 & 4.40 & 4.74 & \textbf{4.17} & 15.57 \\ 
        CubicHermiteSpline & 4.49 & 4.40 & 4.82 & 4.25 & 15.55 \\ 
        \hline
        $L_{paras}$ only & 5.00 & 4.20 & 4.81 & 5.98 & 16.77 \\
        w/o $L_{resample}$ & 4.54 & 4.33 & 4.72 & 4.58 & 15.82 \\ 
        \hline
        $Cobb^s$ only & 5.15 & 4.36 & 6.35 & 4.74 & 16.85 \\ 
        $Cobb^r$ only  & 5.74 & 4.73 & 6.53 & 5.95 & 16.96 \\  
        \hline
        Ours & \textbf{4.32} & \textbf{4.17} & 4.59 & 4.20 & \textbf{15.53} \\ 
        \hline
    \end{tabular}
    \label{tab.ablation-sanyuan}
\end{table}

\begin{table}[!h]
    \caption{Quantitative evaluation of various segmentation models on AASCE2019 dataset. Note in w/o SegRefine and ours, the re-trained SSformer is used for generating the initial segmentation mask. Values of MDC and Seg4Reg+ are reported from \cite{lin2021seg4reg+}. }
    \centering
    \begin{tabular}{lccc}
        \hline
        Method   & IoU & Dice & pixel-AC \\
        \hline
        MDC \cite{Wei_2018_CVPR}    & 75.91    & 86.31     & 96.73  \\
        Seg4Reg+ \cite{lin2021seg4reg+}    & 77.86    & 87.55     & 95.49  \\
        w/o SegRefine    & 91.27    & 95.43     & 98.23  \\
        Ours    & \textbf{93.19}    & \textbf{96.40}     & \textbf{99.24}  \\
        \hline
    \end{tabular}
    \label{tab.segmentation}
\end{table}

\textbf{Effectiveness of the B-spline curve representation.}
In our CurvePred module, we use B-spline curve to fit the initial centerline points.
To verify the benefits of B-spline over other curve representations, we conduct experiments with other representative curve representations for point fitting or interpolation.
Specifically, we perform curve fitting with a 6-degree polynomial curve which is used in \cite{tu2019automatic}, and point interpolation with cubic spline curve and cubic Hermite spline, to the initial centerline points.
The curves obtained by above methods are passed to the AngleEst module to get the final results, while the quantitative results are shown in 3rd to 5th rows of Table \ref{tab.ablation} and \ref{tab.ablation-sanyuan}.
It can be observed our B-spline curve achieves the best performance and the cubic Hermite spline curve is the second best, and the curve fitted by the 6-degree polynomial produces the largest error.
These results verify the B-spline curve is a more suitable spine representation which can support more robust angle estimation.

\textbf{Effectiveness of the loss design.}
To learn the B-spline, we optimize the losses for B-spline parameters and the centerline points (Equation \ref{form.loss}) in the CurvePred module.
We conduct experiments to evaluate the effectiveness for these terms.
First, we train the B-spline prediction model by only using the $L_{paras}$ loss and the results are shown in 6th row in Table \ref{tab.ablation} and \ref{tab.ablation-sanyuan}.
It can be seen that directly predicting the B-spline parameters (i.e., only using $L_{paras}$ loss) leads to inferior angle prediction accuracy. 
In addition to the supervision on B-spline parameters, the centerline points based losses $L_{init}$ and $L_{resample}$ enforce the point-wise supervision for more stable B-spline learning.
To verify their effectiveness, we train a new CurvePred module with $L_{init}$ added to $L_{paras}$, i.e., without $L_{resample}$, and the results are shown in the 7th row of Table \ref{tab.ablation} and \ref{tab.ablation-sanyuan}.
The performance increase comparing to using $L_{paras}$ only indicates explicit point supervision is helpful for B-spline learning.
Moreover, by adding the $L_{resample}$ loss, the performance is further increased in our full model.
Such results verify the effectiveness of the resample loss for providing more constraints for learning the B-spline in a more robust manner.


\textbf{Effectiveness of the hybrid scheme for Cobb angle estimation.}
In the AngleEst module, we combine the curve-slope based angle $Cobb^s$ and the regression-based angle $Cobb^r$ to obtain the final angle.
To evaluate the effectiveness of such hybrid Cobb angle estimation scheme, we compare the results obtained by only using $Cobb^s$ or $Cobb^r$ in the 7th and 8th rows of Table \ref{tab.ablation} and \ref{tab.ablation-sanyuan}.
It can be found the errors of $Cobb^s$ are relatively smaller than $Cobb^r$, indicating the B-spline is well learned and the angles obtained by curvature analysis are fairly consistent with the human annotation.
It is also worthy noting that our final angles obtained by combining $Cobb^s$ and $Cobb^r$ produces the smallest error.
This is due to $Cobb^s$ and $Cobb^r$ may outperform each other in different cases.
By combing them together, we can benefit from the advantages of both angle estimation manners.

\begin{figure}[t]
    \centering
    \includegraphics[width=0.95\linewidth]{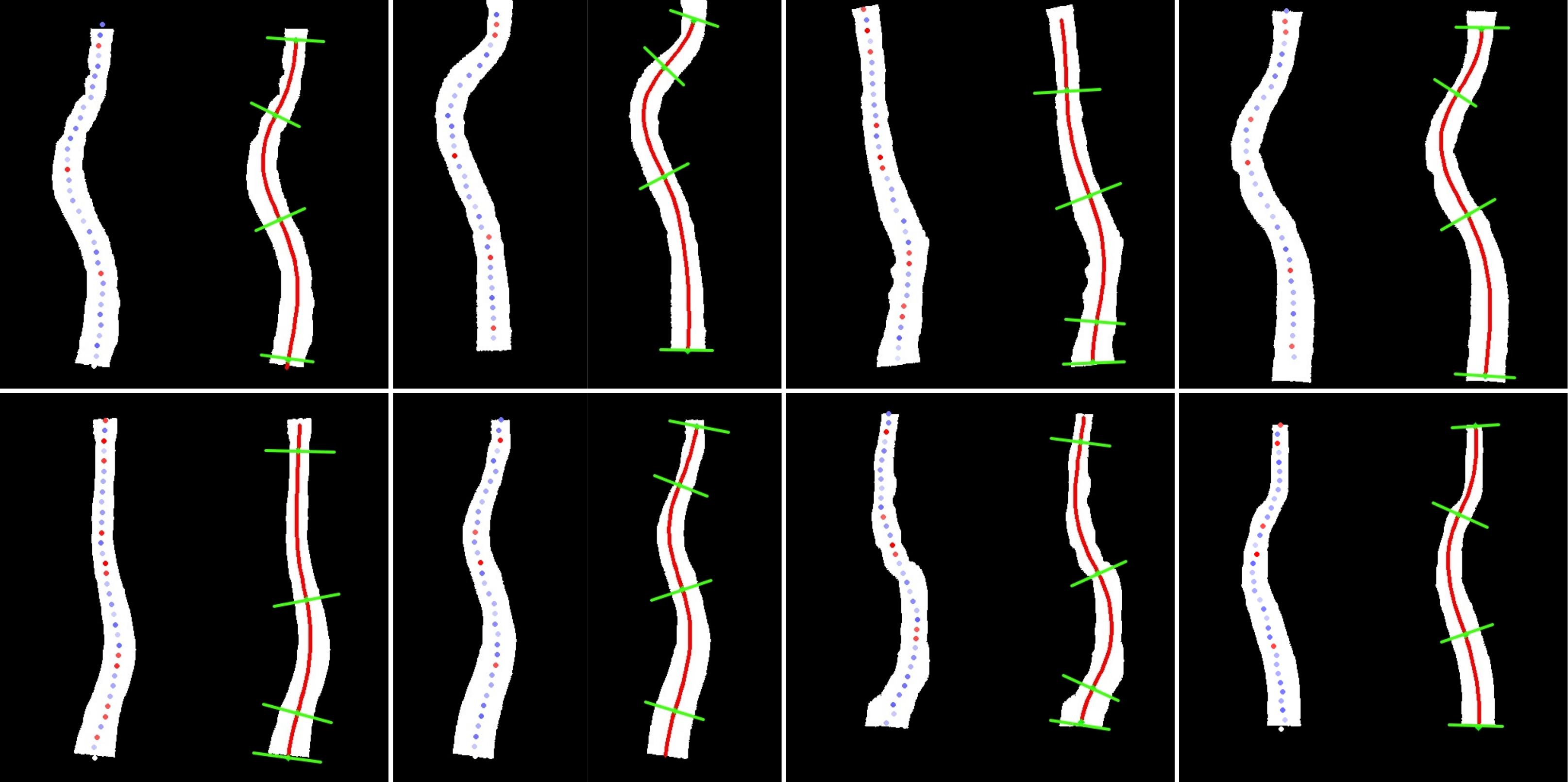}
    \caption{
    Demonstration of the interpretability of the regression model and curve slope analysis for Cobb angle estimation from AASCE2019 (top) and CJUH-JLU (bottom) data.
    The first column of each group shows the point-wise activation visualization with red color represents higher values, and the second column shows the endplates obtained by slope analysis of the B-spline curve.
    }
    \label{fig:points_vis}
\end{figure}

\section{More analysis}

\textbf{Interpretablity of the AngleEst module.}
Our AngleEst module combines the curve slope analysis and the angle regression model to estimate the final Cobb angles.
With the slope analysis, the location of endplates used for measuring the Cobb angles can be directly derived and visualized (see the green segments shown in Figure \ref{fig:points_vis}), which makes the results interpretable.
To further verify the interpretablity of the angle regression model, we follow the idea of Grad-CAM \cite{grad-cam} and visualize the activation or importance of the B-spline sampled centerline points used for predicting the Cobb angles.
Specifically, we choose the second last linear layer of the MLP in the CurvePred module as the target layer and compute its activation map w.r.t to the regression results.
The losses of the predicted angles are backpropagated to the target layer and a 128-dimension vector of gradients is obtained.
Then, we propose a point-wise activation estimation scheme to generate an activation vector for the 34 input centerline points from the 128-dim gradient vector and the 128-dim feature vector of the target layer.
First, we perform element-wise multiplication to combine the gradient and feature vector.
Next, we reshape the 128-dim vector to a 64$\times$2 matrix and compute the mean of each row to get a 64-dim vector.
This step acts similarly as the global average pooling operation as in Grad-CAM \cite{grad-cam}.
Finally, we perform subsampling from the 64-dim vector and get a 34-dim vector as the output activation vector for the 34 input points.
In Figure \ref{fig:points_vis}, we visualize the activation of each point and compare with the visualization of slope-based endplates.
It can be seen the points with higher activation values (shown in red) tend to appear near location with higher curve curvature, demonstrating the prediction of the regression model is also based on interpretable curve features.  



\begin{table}[t]
    \caption{Analysis of $\alpha$ for combining the slope and regression based Cobb angles. Results are obtained on AASCE2019.}
    \centering
    \setlength{\tabcolsep}{1.5pt}
    \begin{tabular}{l|ccc|ccc}
        \hline 
          & \multicolumn{3}{c|}{validation set} & \multicolumn{3}{c}{test set} \\
        \hline
        $\alpha$ & MAE$_{MT}$ & MAE$_{PT}$ & MAE$_{TL}$ & MAE$_{MT}$ & MAE$_{PT}$ & MAE$_{TL}$ \\
        \hline
        0.2 & 3.68 & 5.77 & 5.82 & 3.93 & 4.99 & 5.14\\
        0.3 & 3.58 & 5.48 & 5.34 & 3.81 & 4.73 & 4.88\\
        0.4 & \textbf{3.53} & 5.24 & 4.96 & \textbf{3.73} & 4.72 & 4.74\\
        0.5 & 3.54 & \textbf{4.79} & \textbf{4.66} & 3.76 & \textbf{4.66} & \textbf{4.56}\\
        0.6 & 3.63 & 4.99 & 4.71 & 3.85 & 4.78 & 4.68\\
        \hline
    \end{tabular}
    \label{tab.parameter-analysis-a}
\end{table}

\textbf{The analysis of parameter $\alpha$.}
The main parameter in our pipeline is the weight $\alpha$ for combing the Cobb angles obtained by the curve-slope analysis and curve-based regression.
In our implementation, $\alpha$ is set based on the performance on the validation set.
Also, since each type of Cobb angle (i.e., MT, PT, TL) has different result distribution, $\alpha$ is independently set for each angle type.
Table \ref{tab.parameter-analysis-a} shows the experiment record when we selected $\alpha$ based on the validation set as well as the performance on the test dataset.

\section{Conclusion and discussion}
In this paper, we propose B-Spine, a novel deep learning pipeline for robust and interpretable spinal curvature estimation from low-quality X-ray images.
To our best knowledge, we are the first to represent the spine with a smooth and flexible B-spline curve.
To improve the performance, we propose novel solutions for each module, including the SegRefine network to generate refined mask from the initial segmentation, the multi-step B-spline learning with the resample loss for explicit point supervision and the
hybrid Cobb angle estimation by combining the curve analysis and data-driven approaches.

The superior performance on AASCE2019 and our new proposed CJUH-JLU datasets verifies the effectiveness of our method. Sufficient ablation studies demonstrate the effectiveness of the proposed modules. Since the three modules of our pipeline can be trained independently, it is convenient to improve each module by using latest networks or new architectures. 
Meanwhile, as the later module still depends on the result of former one, the multi-stage pipeline may meet the issue for error accumulation, especially for severe low-quality input.
How to improve the robustness of the pipeline while still keeping the interpretability of different modules is an interesting future work.
Furthermore, although our B-spline learning module is designed for spine centerline approximation, it is also possible to extend the multi-step learning scheme and the resample loss for learning B-spline for other applications.


Benefited from our B-Spine, the workload of the clinicians can be alleviated as the spinal curvature can be estimated in an automatic and interpretable manner.
Also, thanks to our robust pipeline, B-Spine can be applied to low-quality images obtained from the outdated devices and help to improve the medical service for underdeveloped regions.
We believe our B-spline spine representation, the pipeline for annotating and learning the B-spline spine, as well as our CJUH-JLU dataset can inspire more follow-ups for more accurate, more robust and more interpretable Cobb angle estimation.

\section*{Declaration of competing interest}
The authors declare that they have no known competing financial interests or personal relationships that could have appeared to influence the work reported in this paper.

\section*{Acknowledgments}
This work was supported in part by Natural Science Funds of China under Grant 62202199 and Jilin Scientific and Technological Development Program under Grant 20230101071JC.

\section*{Data availability}
Data will be made available on request.

\bibliographystyle{model2-names.bst}\biboptions{authoryear}
\bibliography{references-bib}

\end{document}